\documentclass{sf2a-conf2012}
\usepackage{graphicx}
\usepackage[]{natbib}  
\usepackage[cyr]{aeguill}
\usepackage{epstopdf}
\usepackage{multicol}

\def\BibTeX{{\rm B\kern-.05em{\sc i\kern-.025em b}\kern-.08em
    T\kern-.1667em\lower.7ex\hbox{E}\kern-.125emX}}
\bibpunct{(}{)}{;}{a}{}{,}  

\usepackage{color}

\begin{document}

\TitreGlobal{SF2A 2012}


\title{Toward a model for HFQPOs in Microquasars}

\runningtitle{A model for HFQPOs}

\author{P. Varniere}\address{APC, Universit\'e Paris 7}
\author{M. Tagger}\address{LPC2E, Orl\'eans}
\author{F. Vincent$^1$}
\author{H. Meheut}\address{Physikalisches Institut, Universitat Bern}

\setcounter{page}{237}


\maketitle


\begin{abstract}
There have been a long string of efforts to understand the source of the variability observed in microquasars but no model has yet 
gained wide acceptance, especially concerning the elusive High-Frequency Quasi-Periodic Oscillation (HFQPO). We first list the constraints arising from 
observations and how that translates for an HFQPO model.  Then we present how a model based on having the Rossby Wave Instability (RWI) active in the
disk could answer those constraints.\end{abstract}



\section{What does a HFQPO model need to explain}
  
       Even if High-Frequency Quasi-Periodic Oscillations are much weaker than their Low-Frequency counterparts we now have data from several outbursts from
       eight different sources. Indeed, sources like XTE J$1550$-$564$, have {exhibited HFQPOs} with enough regularity to obtain a stringent 
        list of constraints for any theoretical model wishing to provide an explanation for them [3].
 \newline 
 
        $\bullet$ The first observational fact that one needs to explain is the {\bf modulation of the flux} associated with the frequency. Indeed, even if HFQPO has a rms amplitude much 
        lower than in the case of the {LFQPO, the flux still modulates} at a level of a few percent and it {has been} shown to be stronger at higher energies 
        (see [3] for examples). 
 \begin{figure}[ht!]
 \centering
 \includegraphics[width=0.42\textwidth,clip]{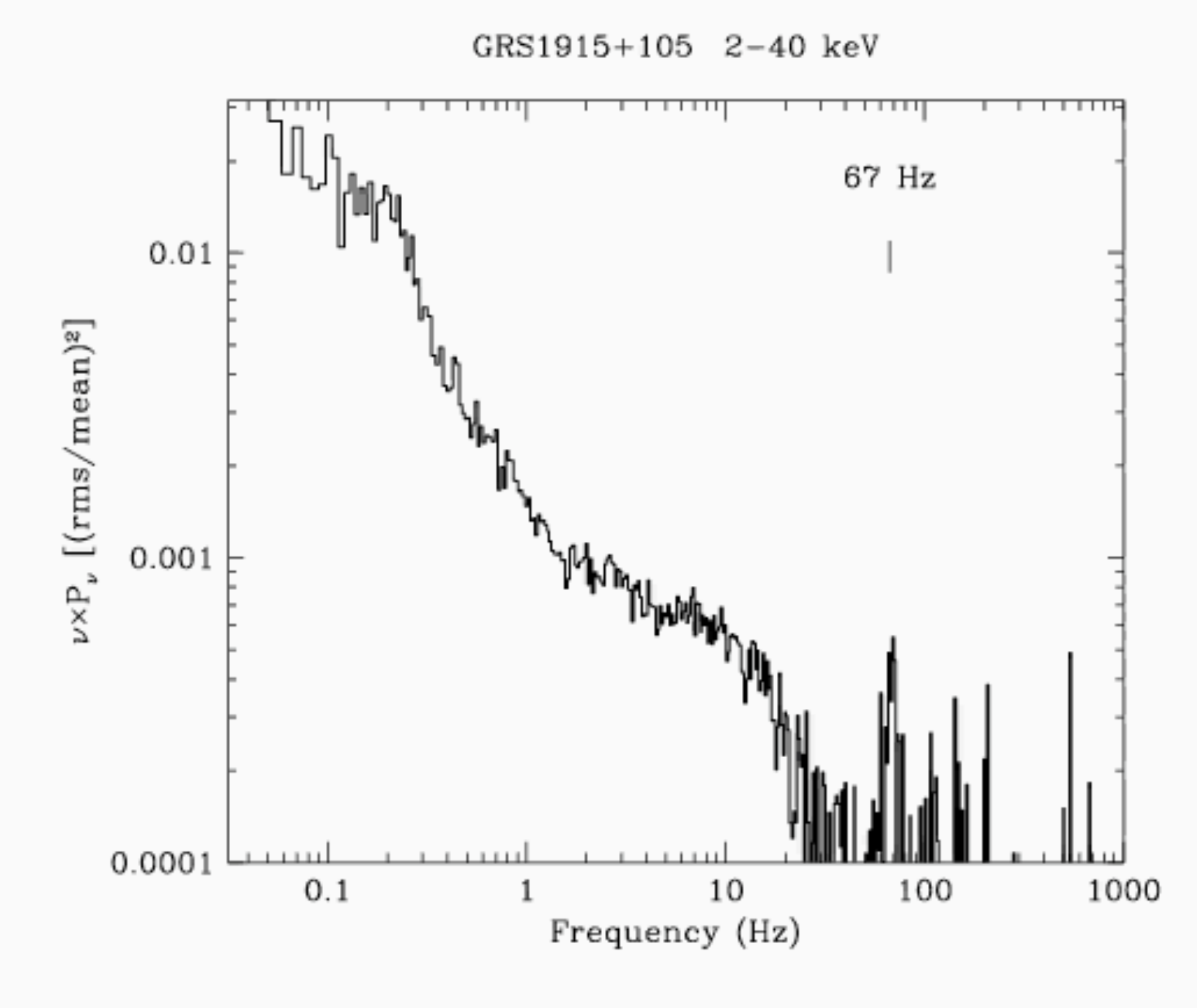}%
 \includegraphics[width=0.44\textwidth,clip]{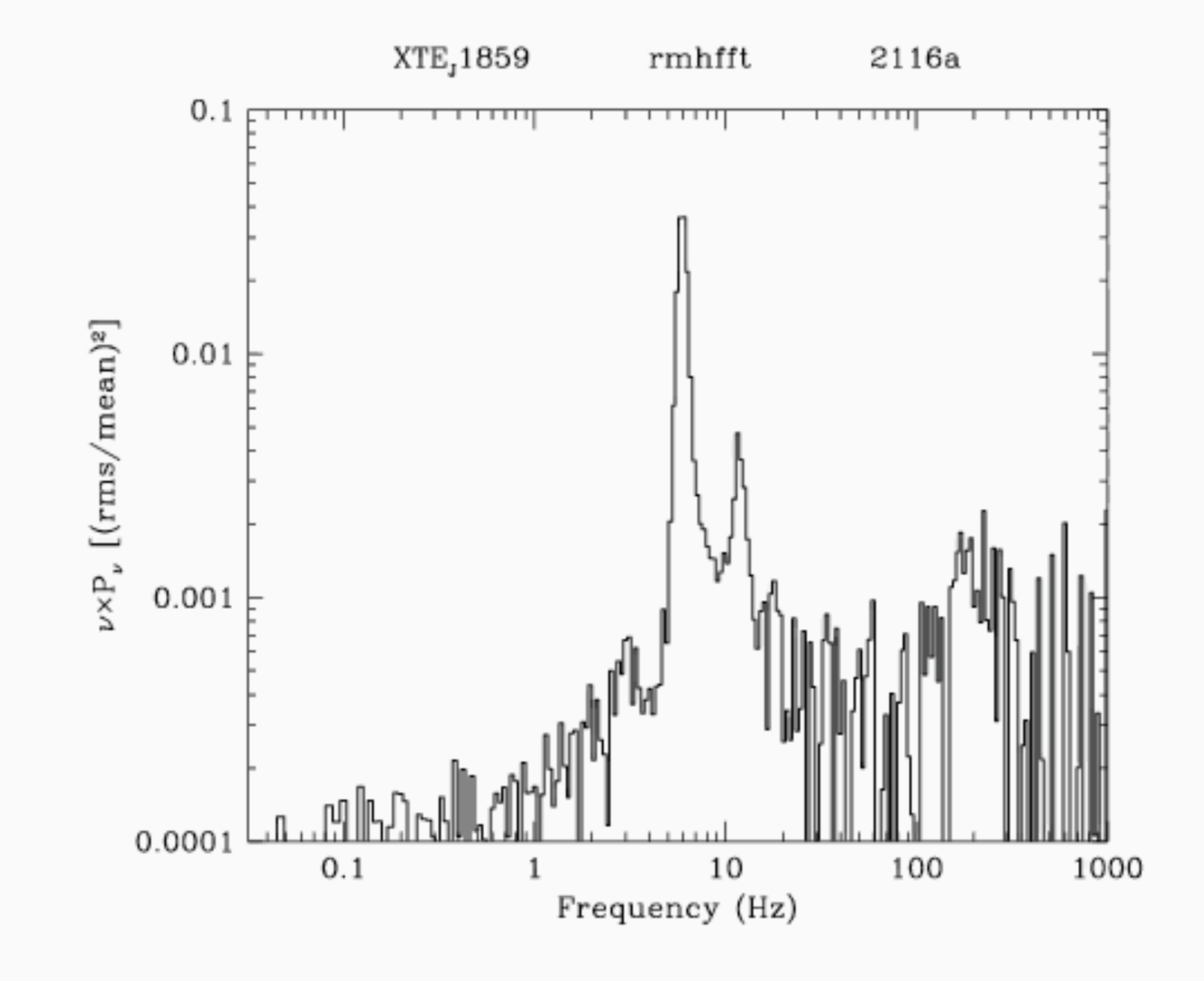}      
  \caption{{\bf Left:} Power density spectrum (PDS) of GRS 1915+105 showing  the $67$Hz. {\bf Right:} PDS of XTE J1859+226 with LFQPO and HFQPO. }
  \label{varniere:fig1}
\end{figure}
       $\bullet$ Since the observation of the $67$Hz HFQPO of GRS 1915+105 we know that HFQPOs can occur in the absence of LFQPOs. It is therefore required
       that the HFQPO model be independent of the LFQPO model. However, most HFQPO detections do occur in the presence of a LFQPO, 
       as observed for example during the outbursts of XTE J1550-564 or XTE J1859+226.
       When they co-exist we have type A and B LFQPOs, not the standard type C. 
       
       All of this demonstrates that, even if {\bf HFQPO and LFQPO models need to be independent,
       they also need to be coherent with each other as the two QPOs co-exist in the disk}. This is a more stringent requirement than it may at first seem, as one need not only 
       find a model for the HFQPO's characteristics, but also a model that can coexist in a disk with an LFQPO.

\begin{multicols}{2}

            $\bullet$ Another exacting  requirement coming from observation is the fact that the  {frequencies of HFQPOs, albeit more} stable than in the case of the LFQPO, 
      show a small but significant variation. In the case of XTE J$1550$-$564$ the figure on the right represents the observed occurrences of the HFQPOs in $10$Hz bins
      during the outburst of 1998-99 and 2001.       \\
       {\bf Any model aiming to explain the HFQPOs {must} be able to reproduce the observed dispersion in the frequency.}
      
\hspace*{-0cm} \includegraphics[width=5.5cm]{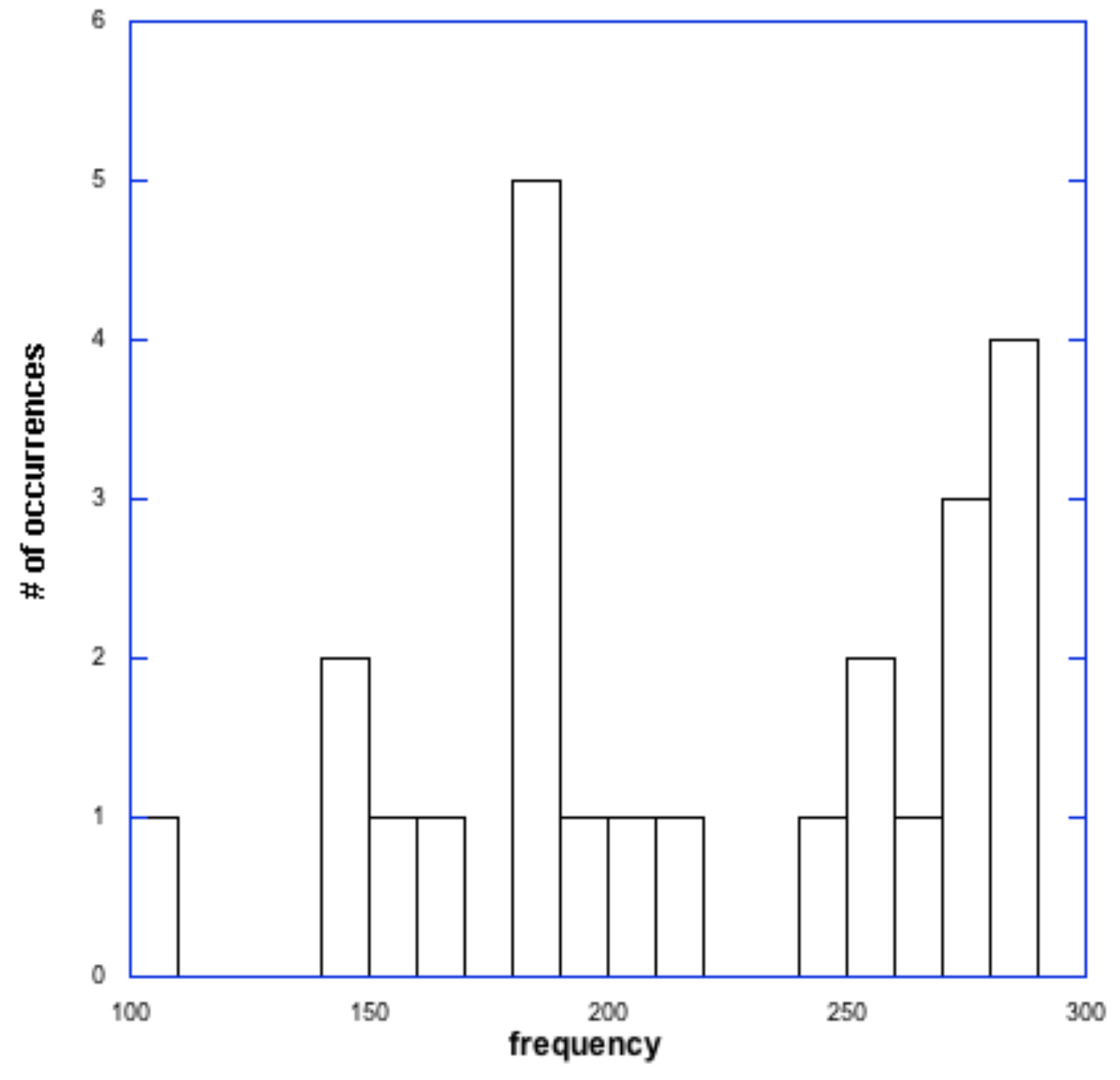}      
 \end{multicols}

      \vspace*{-0.5cm}$\bullet$  Another {characteristic of HFQPOs} is that they can be observed either alone {\bf or} in \lq pairs\rq\  (with closely related frequencies, 
      most of the time near a 2:3 ratio). {\bf This {points toward} a mechanism that can select several linked frequencies depending on the disk conditions.} \\

      Any model {wishing} to explain {HFQPOs must} be able to explain this small but {stringent list of requirements}. As we get more observations with future 
      {detectors} we will be able to add to this list and further {constrain} the models.

\section{The   {Rossby Wave Instability}  as a model for HFQPOs}

The   {Rossby Wave Instability} is an hydrodynamical instability that occurs {in the presence} of an extremum of the 
 vortensity (defined by $\Sigma\Omega/(2\kappa^2)\cdot p/\sigma^\gamma$ where 
 $\kappa$ is the disk epicyclic frequency, $\Omega$ {is the} rotation frequency and $\sigma$ {is the} surface density).
 Because of its {characteristics}, we proposed the RWI as a possible explanation for HFQPOs [4]. \\

 \begin{figure}[ht!]
 \centering
 \includegraphics[width=0.4\textwidth,clip]{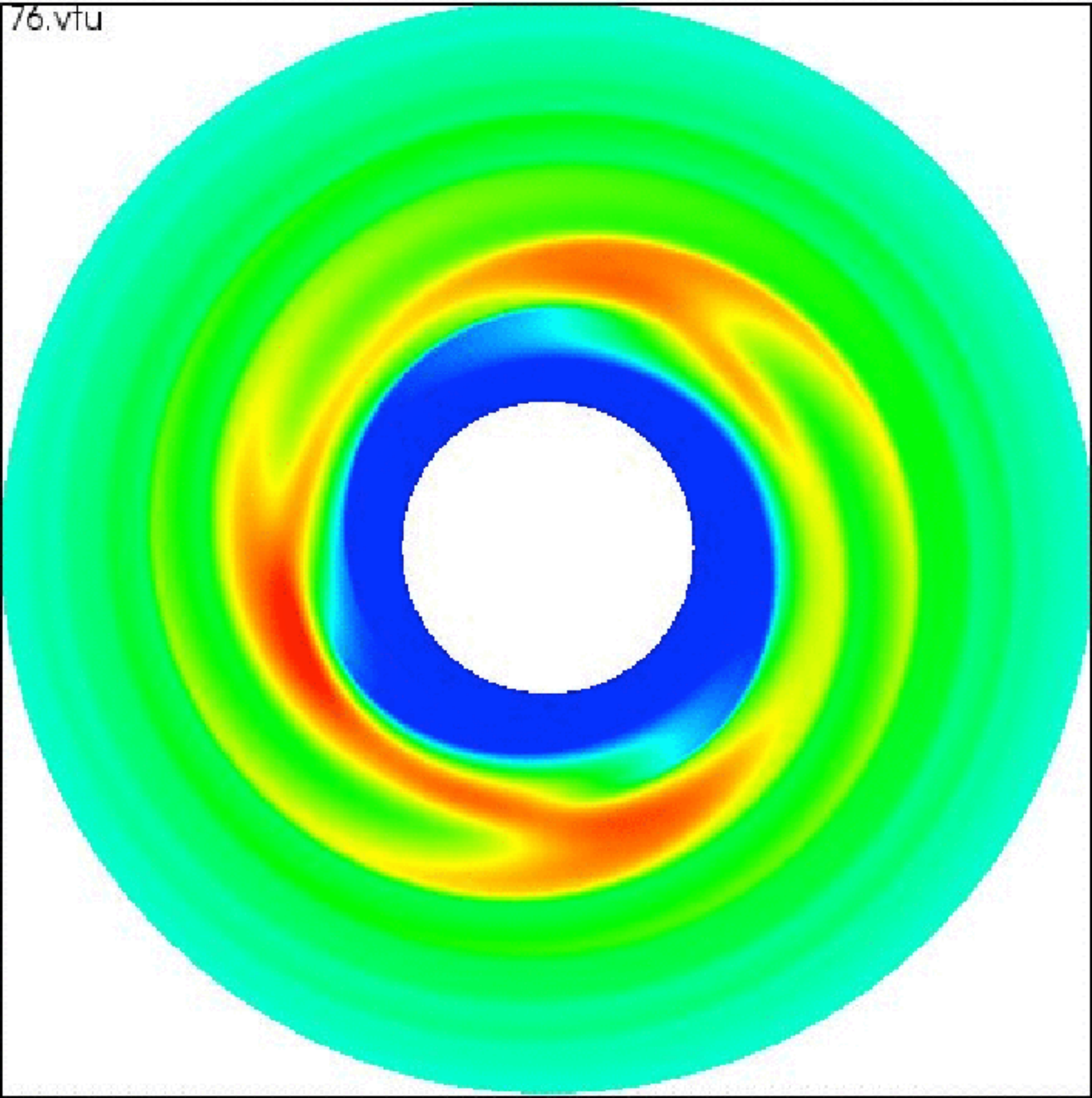}%
 \includegraphics[width=0.4\textwidth,clip]{{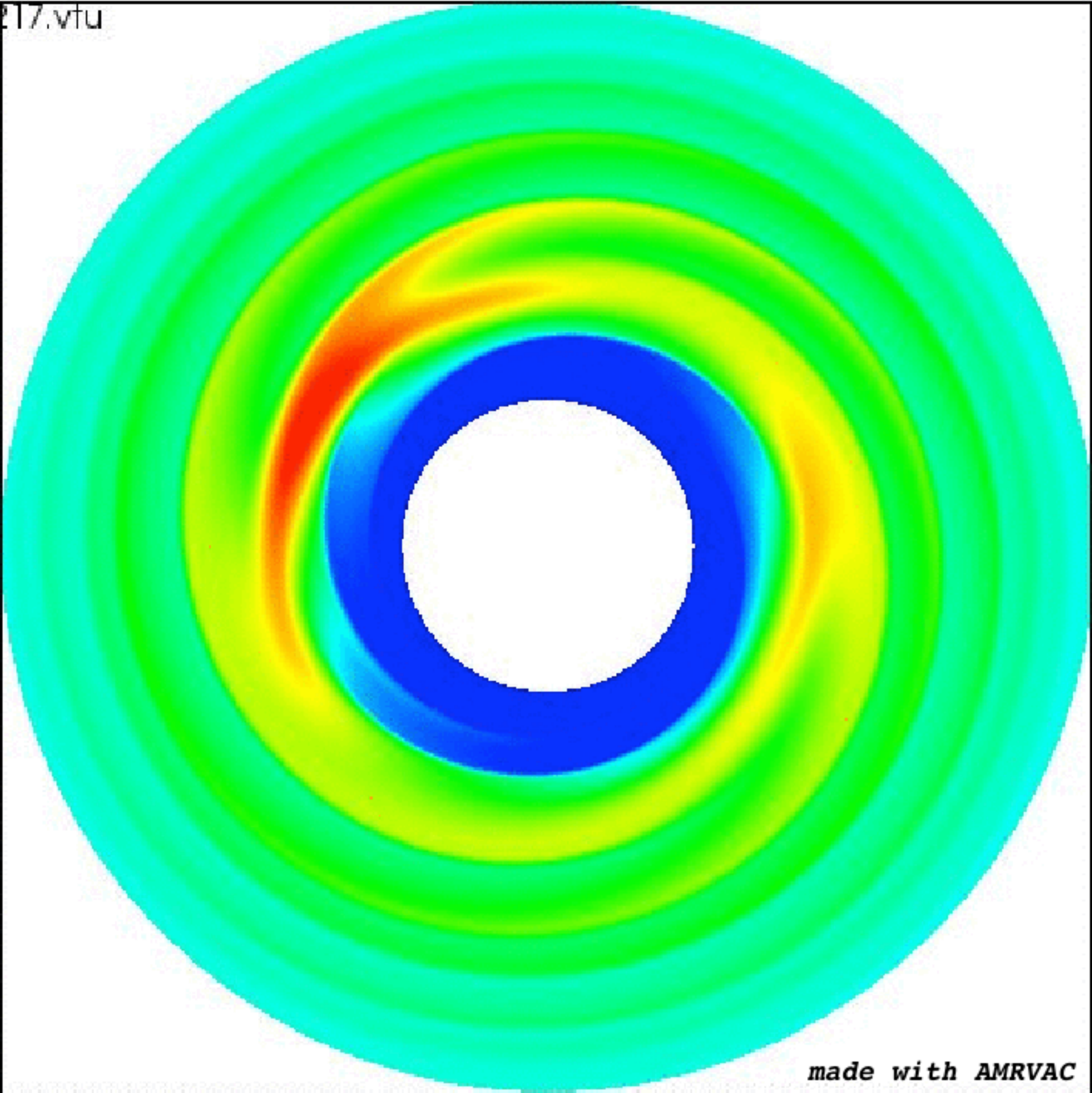}}      
  \label{varniere:fig2}
    \caption{ Density slab of the disk from an hydrodynamic simulation of the RWI at two different {times}, using the code AMRVAC}
\end{figure}

 {$\bullet$} \   In the case of a disk {in which the inner edge approaches} its last stable orbit an extremum of the vortensity becomes possible therefore leading to the RWI
 as shown in the hydrodynamic simulation shown on Fig.\ref{varniere:fig2}.  
  {These graphs} represent a slab ($z=0$) of  the density in a $3$D disk in the Paczynsky \& Wiita (1980) gravitational potential, which means a spin $a=0$. 
   We {later} used modified Newtonian potential [1] to model the full range of spin and confirmed the results.
    
    Because the inner edge of the disk {must} be close to its last stable orbit but not \lq exactly\rq\  at it, there is a  small {radial range}
    where the instability can develop [4] leading to a change in observed frequency. As {we do not know} precisely the density profile in the disk, especially
    close to its last stable orbit, it is hard to put a hard boundary on the frequency changes but it could reach $30$\% without a dramatic change to the {profiles}. \\

     {$\bullet$} \    Another interesting point is that the RWI does not {require} the disk to be in the condition {for a} LFQPO model to occur [4].
    Nevertheless, the RWI was also demonstrated to be stronger {in the presence} of a vertical magnetic field[4] 
    and we have recently shown the ability of the RWI and the AEI (a candidate to explain the LFQPO) to co-exist in a magnetized accretion disk  [5,6].  
    Therefore, it could give rise to either {HFQPOs alone} or HFQPOs {and LFQPOs} depending on the disk condition, as is observed.\\

 {$\bullet$} \  From numerical simulation, we also found that, depending on the disk {conditions,} the dominant mode {can be $m=3, m=2$,} (see Fig.\ref{varniere:fig2})
 {more rarely $m=1$,}  or a mix of {these[4]} which {fit well} with the observed {characteristics of HFQPOs.} \\

\begin{figure}[ht!]
 \centering
 \includegraphics[width=7cm]{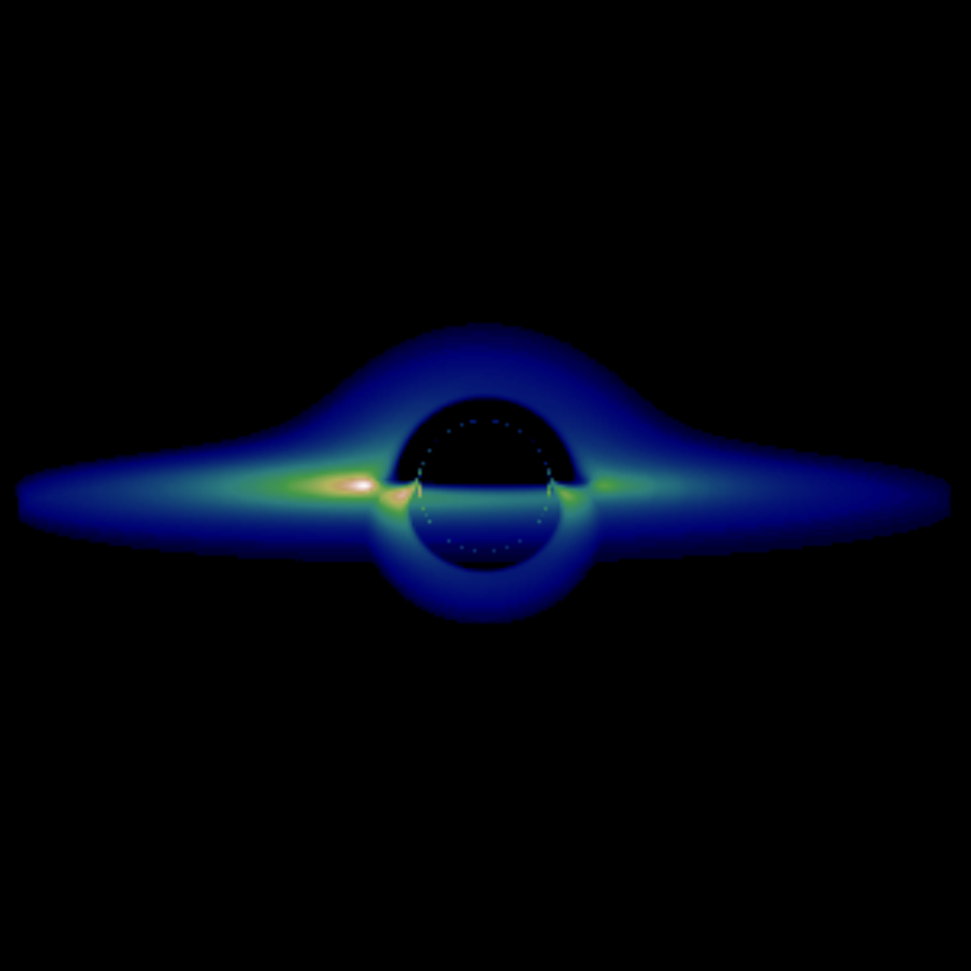} 
 \includegraphics[width=7cm]{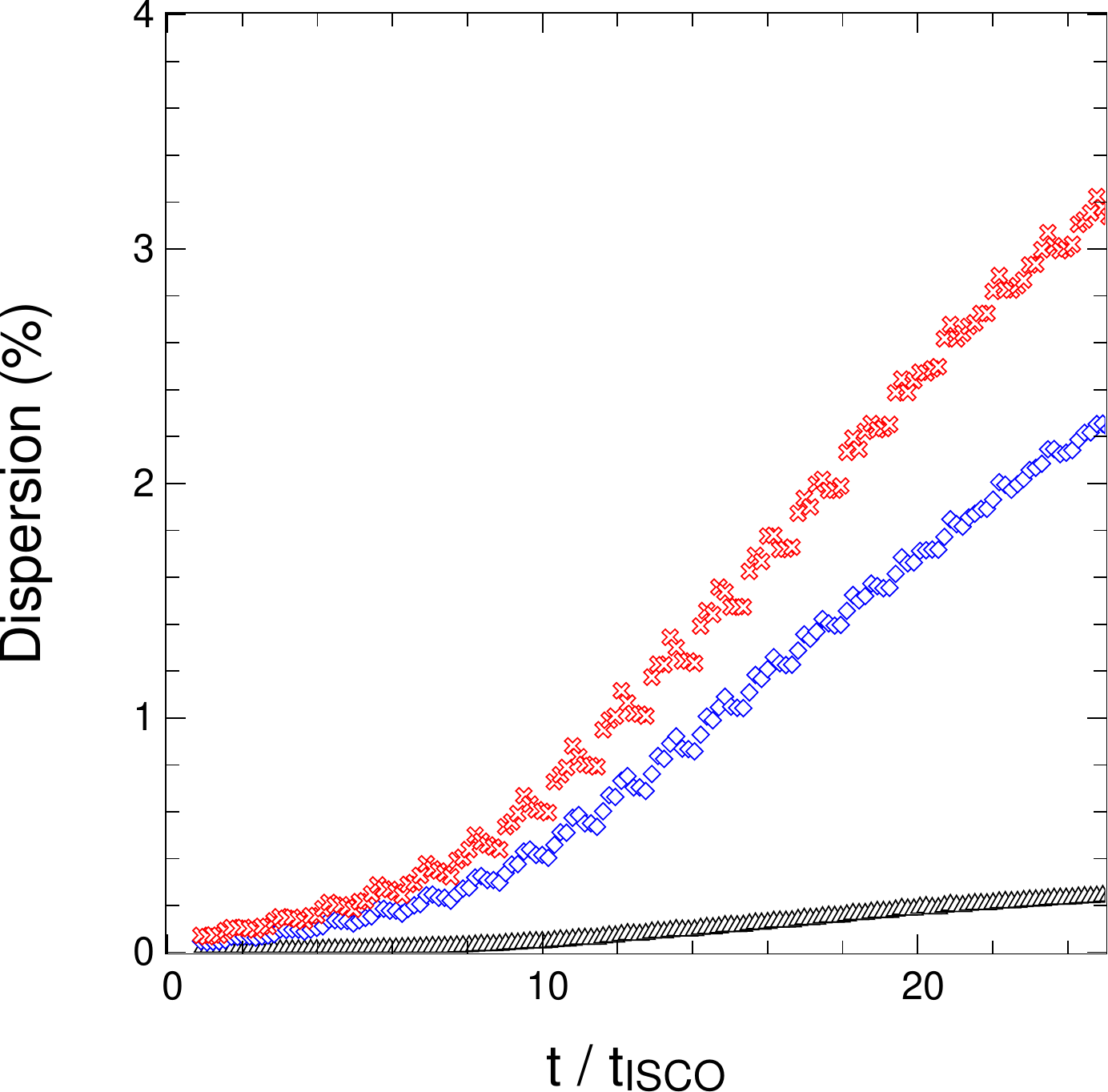}
  \caption{ Left: Ray tracing of a 3D simulations of the RWI at 85$^\circ$  inclination. Right: time evolution of the rms of the flux modulation at 85$^\circ$ , 45$^\circ$  and 5$^\circ$ inclination.}
  \label{varniere:fig3}
\end{figure}

 {$\bullet$} \  We now perform 3D {simulations} of the RWI and confirm the previous $2$D and analytical results and also {produce} the associated 
 image(see Fig.\ref{varniere:fig3})/light curve using the code Gyoto [7]. 

When the RWI is active in the disk the light curve is modulated up to a few \% [8] and this modulation is energy dependent. 
The precise impact of the spin, especially the case of high spin, is still under study, but the RWI is present and modulate the flux

\section{Conclusions}
The RWI is a promising model {for HFQPOs} as it {gives} rise to {several} observed features such as the possibility to have 
small {variations} in the frequencies as well {as mode} selection depending on the  {conditions} in the disk. Moreover, this instability can co-exist with the
Accretion-Ejection Instability [5,6] proposed as a model for the ubiquitous LFQPO. Lastly, we have recently {shown} that this instability can
effectively modulate the X-ray flux within the observed limit [8]. {In the future,} we {will explore} the impact of the spin of the black hole, the link {with ejection} and 
the overall evolution of the system.

\begin{acknowledgements}
This work has been financially supported by the GdR PCHE in France and the "campus spatial Paris Diderot". 
CINES?
\end{acknowledgements}


\bibliographystyle{aa}  
\bibliography{sf2a-template} 

\end{document}